# Generating Visual Information for Motion Sickness Reduction Using a Computational Model Based on SVC Theory

Yujiro Tamura, Takahiro Wada, Member, IEEE, Hailong Liu, Member, IEEE

*Abstract*—With the advancements in automated driving, there is concern that motion sickness will increase as non-driving-related tasks increase. Therefore, techniques to reduce motion sickness have drawn much attention. Research studies have attempted to estimate motion sickness using computational models for controlling it. Among them, a computational model for estimating motion sickness incidence (MSI) with visual information as input based on subjective vertical conflict theories was developed. In addition, some studies attempt to mitigate motion sickness by controlling visual information. In particular, it has been confirmed that motion sickness is suppressed by matching head movement and visual information. However, there has been no research on optimal visual information control that suppresses motion sickness in vehicles by utilizing mathematical models. We, therefore, propose a method for generating optimal visual information to suppress motion sickness caused from vehicle motion by utilizing a motion sickness model with vestibular and visual inputs. To confirm the effectiveness of the proposed method, we investigated changes in the motion sickness experienced by the participants according to the visual information displayed on the head-mounted display. The experimental results suggested that the proposed method mitigates the motion sickness of the participants.

## I. INTRODUCTION

With the advancement of automated driving, there is concern regarding an increase in motion sickness[1][2][3]. This concern arises because of design changes in the vehicle interior, the shift from driver involvement to a more passive role, and increased engagement in in-vehicle visual tasks, such as reading or working on a laptop PC. Consequently, the mitigation of motion sickness is crucial for the progress and development of automated vehicles (AVs).

Computational or mathematical models of motion sickness have been actively developed to estimate the severity or symptom because they are useful for deriving countermeasures. The motion sickness dose value (MSDV), defined as the frequency-weighted sum of acceleration, is a widely used fitting model [4]. Models based on the mechanism of motion sickness have also been developed. Oman [5] developed the first computational model based on the hypothesis of motion sickness, in this case, the sensory conflict or neural mismatch theory[6]. This has triggered research into the hypothesis of motion sickness using computational models. The subjective vertical conflict (SVC) theory [7] postulates that motion sickness arises from the cumulation of the difference between the gravitational vertical sensed by the sensory organs and the estimation made by the internal models in the central nervous system (CNS). A computational model of the SVC theory was developed to describe motion sickness during one degree-of-freedom (DoF) vertical motion [8]. Subsequently, several models of the SVC theory have been developed, including multi-DoF motion[9][10], vestibular-visual interaction[11][12][13], and the motion prediction effect[14]. Several efforts have been made to apply computational models to derive countermeasures. For example, in the context of AV control, there are methods for acceleration pattern generation [3][15] and trajectory generation on the road [16][17]. The model has also been adopted for vehicle-lean control [18], mainly for manually controlled vehicles. All the above solutions mitigate sickness based on vehicle motion.

There are situations where the need arises for strategies to mitigate motion sickness without altering or controlling vehicle motion. This becomes particularly significant when the vehicle motion cannot be adjusted according to individual preferences, as is the case in public transportation, or when a more cost-effective solution is required. The possibility exists to manipulate visual information. In fact, countermeasures to the sickness in reading a book or watching a smartphone while riding a car were proposed. For example, research has attempted to reduce motion sickness when reading characters on a smartphone by moving spheres drawn by computer graphics around the text in accordance with the vehicle's motion [19]. In addition, research studies have proposed methods to mitigate sickness when reading a book on a head-mounted display (HMD) while riding a moving machine by fixing the book to the Earth-fixed coordinate system [20] and by displaying the surrounding images using a video see-through technique[21]. However, to date, no research has been conducted on the optimization of visual information manipulation to minimize motion sickness, although models that can handle both vestibular and visual information have been developed [12][13].

Therefore, the purpose of the present study is to propose a method that utilizes a model from [9] to generate optimal visual information to mitigate motion sickness during vehicle or head motions. An experiment was conducted with participants wearing HMDs exposed to longitudinal fore-aft motion in an automated personal vehicle. This study examined the effectiveness of the proposed method, which generates visual inputs displayed on a Head-Mounted Display (HMD), in reducing motion sickness.

## II. GENERATION METHOD OF VISUAL INFORMATION USING A MODEL OF MOTION SICKNESS

### A. Motion sickness model with vestibular-visual interaction

Fig. 1 shows a computational model introduced in the present study for predicting motion sickness using vestibular and visual signals, based on the model introduced in [12]. This model is based on the SVC theory [7], which postulates that the difference between the direction of gravity or verticality sensed by the sensory organs and that expected using the internal model of sensory dynamics provokes motion sickness. The model output is the motion sickness incidence (MSI), which refers to the percentage of individuals who experience emesis under specific conditions.

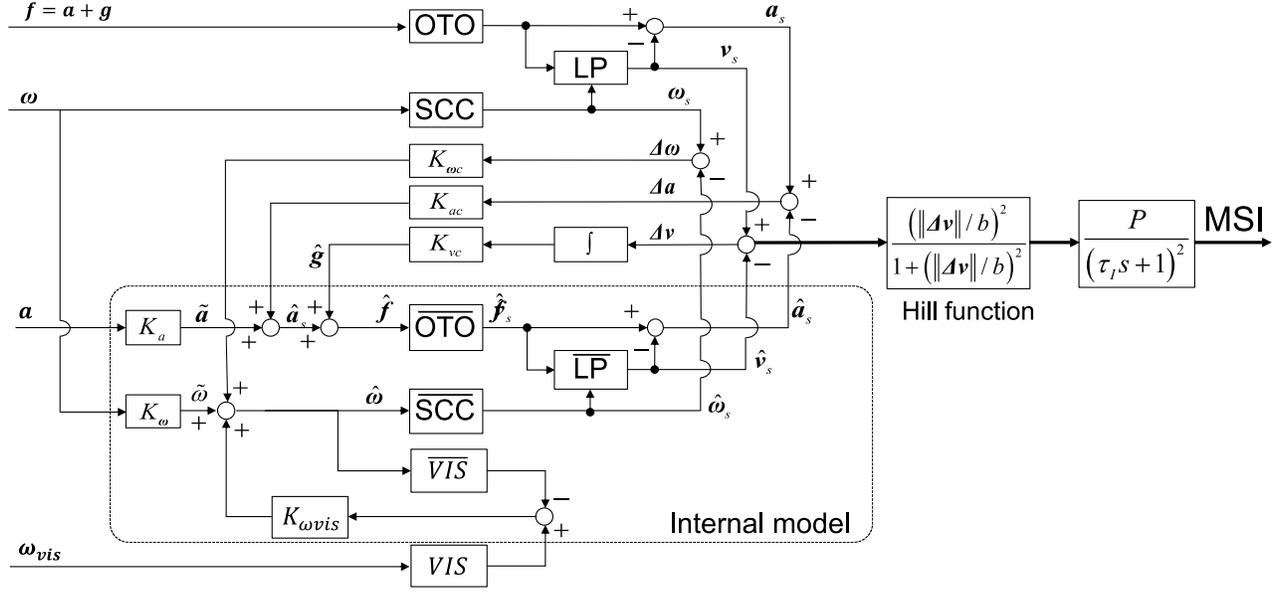

Figure 1. Motion sickness model with vestibular-visual inputs

The upper part of the model represents the sensory dynamics of the vestibular system, which includes the semi-circular canal (SCC) and otolith (OTO). The unit matrix is used for OTO, and it receives the gravito-inertial acceleration (GIA), which is defined as $f=g+a$, where $g$ and $a$ denote the gravitational and inertial accelerations, respectively. The dynamics of the semi-circular canals in transfer function form are given by (1) [22]:

$$\omega_s = \frac{\tau_d s}{\tau_d s + 1} \omega \quad (1)$$

where $\tau_d$ is a time constant.

The neural representation of the vertical direction $v_s$ is assumed to be obtained from the GIA $f$ using (2).

$$\frac{dv_s}{dt} = \frac{1}{\tau}(f - v_s) - \omega_s \times v_s \quad (2)$$

which corresponds to the block LP in Fig. 1.

The lower section of the block diagram is involved in the processing of visual sensory information. The vector $\omega_{vis}$ denotes the angular velocity perceived by the visual system, which is calculated from the optical flow analysis of the video image as in [9].

The angular velocity perceived by the visual system, denoted as vector $\omega_{vis}$, is calculated by analyzing the optical flow of the video image, following a similar approach as described in [12]. This vector serves as the input for the VIS block, representing a part of the process of visual perception. In the present study a unit matrix was used for VIS.

The internal models for both the visual and vestibular systems can be found in the central portion of the diagram. The internal models are labeled by adding a superscript bar to the names of the corresponding sensory dynamics. The dynamics of internal models are assumed to be identical to the corresponding sensory dynamics. For instance, $\overline{SCC}$ is an internal model of the $SCC$ block, and $\overline{SCC} = SCC$ holds.

Humans are assumed to increase accuracy in estimating their body movement by efference copies and/or somatic sensations. The gains $K_\omega$ and $K_a$ describe the accuracy of the angular velocity and acceleration obtained from such estimation. Variables $\Delta a$, $\Delta v$, $\Delta \omega$ are defined as the discrepancies between sensory afferents and those expected by the internal models. These discrepancies are thought to decrease through the feedback process with the gains $K_a$, $K_v$, and $K_\omega$. Note that integration is included only for $\Delta v$, but not for $\Delta a$, $\Delta \omega$, as in [23], whereas two integrals were used for $\Delta g$, $\Delta a$ in [11], which is the original SVC model that can deal with vestibular-visual interaction. To supplement these processes, a feedback mechanism that incorporates VIS, denoting self-motion perception through vision, and $\overline{VIS}$, its associated internal model, was used to adjust the motion perception signals. Finally, the predicted MSI was computed based on $\Delta v$, which represents the discrepancy between the sensed and anticipated vertical signals using a Hill function and a second-order lag element. The model parameters utilized in this study are outlined in TABLE I. The parameter $K_{\omega vis}$ which is related to visual input was taken from [12], and all other parameters were taken from the state-of-the-art model of vestibular motion sickness [23].

TABLE I. PARAMETERS USED IN THE MODEL

| $K_a$ | $K_\omega$ | $K_{\omega c}$ | $K_{vc}$ | $K_{ac}$ |
|---|---|---|---|---|
| 0.1 | 0.8 | 10 | 5.0 | 1.0 |
| $K_{\omega vis}$ | $\tau_d$ [s] | $b$ [m/s$^2$] | $\tau_I$ [min] | $P$ [%] |
| 10 | 7.0 | 0.5 | 12 | 85 |

### B. Generation method of visual stimulus to reduce motion sickness

Assume that head movement in a certain time horizon $f(t+i)$ and $\omega(t+i)$ ($i=1, \ldots, N$) can be obtained, for example, in AVs. In such situations, we propose a method for generating optimal visual information to minimize motion sickness, as shown in (3).

$$[\omega_{vis}^{t+1|t}, \omega_{vis}^{t+2|t}, \cdots \omega_{vis}^{t+N|t}] := \arg\min \sum_{i=1}^{N} L(t+i) \quad (3)$$

where $L(t)$ denotes a stage cost. The candidates include $L(t) = MSI(t)$, which denotes the MSI at time $t$ predicted by the proposed model by inputting the predicted head motion $f(t+i)$ and $\omega(t+i)$. Note that utilizing vehicle motion as a substitute for head motion or developing a model to predict head motion based on vehicle motion are potential approaches for acquiring future head motion.

In the present study, repeated fore-aft motion was employed for vehicle motion for the sake of simplicity. Thus, the optimal visual input signal was calculated more simply, as follows: First, the optimal visual input is expressed by a nonlinear regression of the acceleration as follows:

$$\omega_{vis}(t) := h_0 + \sum_{i=1}^{N} h_i \tan^{-1} a^i(t) \quad (4)$$

where $a(t)$ is the longitudinal inertial acceleration. The coefficients $h_i$ were identified by solving the following optimization problem in preliminary experiments:

$$[h_0, h_1, \cdots, h_N] := \arg\min MSI(T) \quad (5)$$

where $T$ denotes the time when the motion stimulus is terminated. Based on this regression, the visual information to be displayed to the participant can be generated by inputting the current head acceleration at each time instance $t$ without predicting future motion. Note that the vehicle acceleration, measured by the motion tracker, was utilized as a substitute for head motion to calculate (4). $N=10$ was used in this study.

### D. Numerical simulation of proposed method

A numerical simulation was conducted to understand the effect of visual information generated by the proposed method on motion sickness. In the simulation, the longitudinal inertial acceleration of the head (or vehicle) is determined using (6).

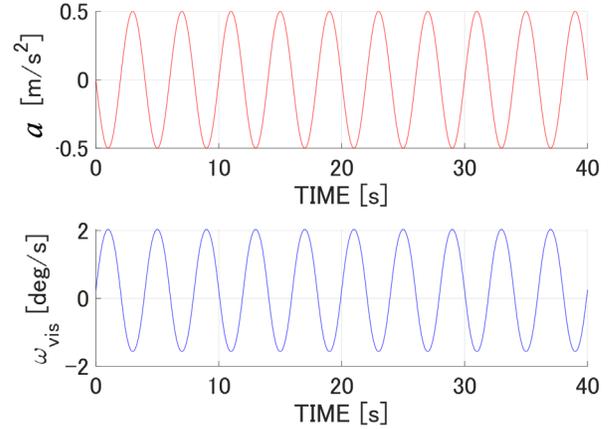
Figure 2. Visual angular velocity generated by the proposed method

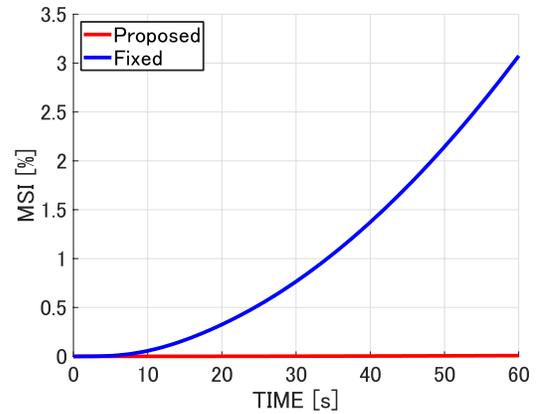
Figure 3. Predicted MSI in numerical simulation

$$a(t) := A \sin 2\pi f t \quad (6)$$

where $A=0.5$ m/s$^2$ and $f=0.25$ Hz were used. By inputting GIA $f(t)=[a(t), 0, g]^T$ into the proposed method, the visual angular motion in the pitching direction shown in Fig. 2 was generated, where the scalar $g$ denotes gravity acceleration.

Fig. 3 shows the MSI predicted by the proposed model by inputting the visual information generated by the proposed method. For comparison, the calculated MSI when $\omega_{vis} = 0$ was inputted as the visual information to the model, is also shown in Fig. 3. The calculated MSI was dramatically reduced by the visual information obtained using the proposed method.

## III. EXPERIMENTAL METHOD

### A. Design

To investigate the effectiveness of the proposed method, experiments with participants were conducted. The motion sickness of the participants exposed to the fore-aft motion of an automated personal mobility vehicle and the visual stimulus provided by an HMD was observed. The visual motion stimulus was provided by the motion of randomly aligned spheres displayed through the HMD. The visual stimulus factor was set as an independent variable consisting of two levels 1) proposed and 2) fixed conditions. In the proposed condition, the visual angular motion in the pitch direction was generated, by the proposed method whereas in the fixed condition, the spheres did not move. The visual stimulus factor was treated as a within-subjects factor; in other words, each participant was exposed to both conditions on different days. The present study received approval from the Ethics Review Committee of Nara Institute of Science and Technology (#2021-I-38).

*B. Apparatus*

Fig. 4 shows the apparatus used in the experiment. An automated version of the small personal mobility vehicle was employed to create a motion stimulus. The visual motion was presented to the participants using an HMD VIVE PRO EYE HMD. A neck brace was used to minimize the neck motion while riding. Vehicular acceleration was measured using a motion tracker at 30 Hz. This signal was used to generate a visual stimulus using (4). The angular velocity and acceleration of the participants' head and eye movements were measured at 30 Hz using sensors mounted on an HMD.

Fig. 5 illustrates a method how the visual stimulus was provided to participants. In a virtual environment displayed through the HMD, an invisible sphere with a radius of one meter was located around the participant's eye. The sphere fixed to the head-fixed coordinate system. Small visible spheres were randomly located on the invisible sphere. Fig. 6 shows the visible spheres seen from the HMD. These visible spheres rotate uniformly around the participant' eyes to generate visual stimulus $\omega_{vis}$. In the fixed condition, the angular velocity of the visible spheres is zero. In the fixed condition, spheres did not move at all.

*C. Participants*

Seven males in their 20s who reported that they did not have any vestibular diseases participated in this experiment. The susceptibilities of the participants to motion sickness, measured by a short version of the motion sickness susceptibility questionnaire (MSSQ)[24], ranged from 0% to 73.5%, with a mean of 43.1% and a standard deviation of 25.4%.

*D. Procedure*

Initially, the participants were explained the experimental methods and procedures and provided written informed consent. Each participant sat in the vehicle in an upright posture and wore an HMD. The participant was exposed to 15 min of fore-aft linear movement of the vehicle with either the proposed or fixed condition. The participants orally reported the symptoms of motion sickness every minute using the MIsery SCale (MISC) [25], in which an 11-point Likert scale was used to subjectively assess the symptoms of motion sickness. The motion of the vehicle was terminated when the reported MISC was equal to or greater than five. After completing the experiment, the participants kept sitting in the stationary vehicle to observe recovery from the sickness for 5 min, during which MISC was also reported every minute. Participants were exposed to each condition of the visual stimulus factor on a separate day. Four of the seven participants were exposed to a fixed condition on the first day, whereas the remaining participants experienced the proposed condition on the first day.

IV. RESULT

Fig. 7 illustrates the temporal progression of MISC in each of the seven participants. As shown in the figure, in the proposed method, the maximum MISC value was smaller or the maximum value was reached earlier than in the fixed condition for six of the seven participants.

Fig. 8 shows the mean of MISC over time from the motion onset to motion termination for each participant. For participants whose motion exposure was terminated, the observed MISC value up to the end time was analyzed in both the proposed and fixed conditions. Based on the visual inspection, the motion sickness was smaller with the proposed method compared to the fixed condition though no statistical test was not conducted since the number of participants was small. This demonstrated that the proposed method has an effect to reduce the motion sickness by changing visual information though the vehicle motion was identical.

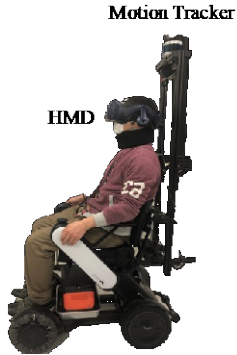

Figure 4. Apparatus

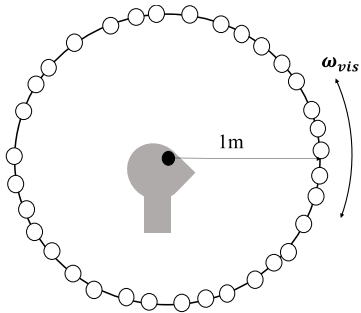

Figure 5. Virtual environment to display visual stimuls

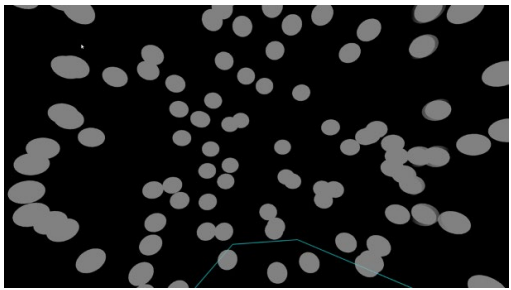

Figure 6. Visual stimulus on HMD

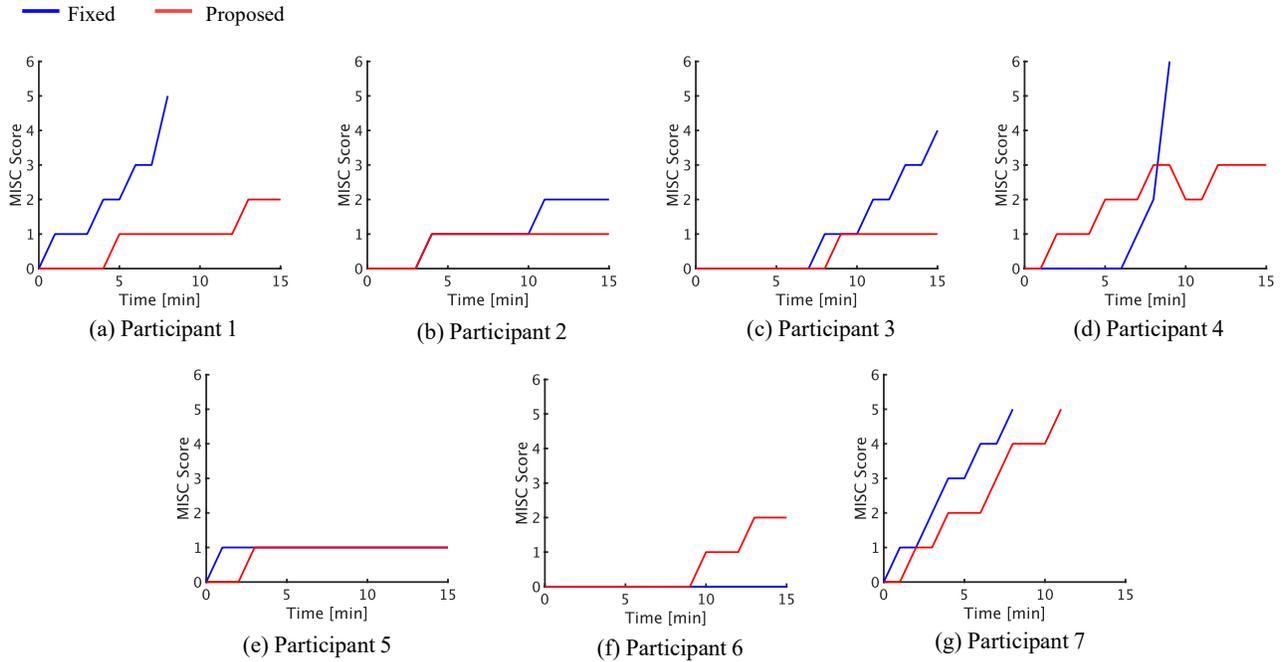

Figure 7. Observed motion sickness for each paticipant

## V. DISCUSSION

The numerical simulation demonstrated that the estimated MSI was reduced when visual information generated by the proposed method was provided to humans, compared to the case with visual information congruent with head motion. Furthermore, the experiment with participants also demonstrated that motion sickness was suppressed or the onset of sickness was delayed in six of the seven participants. Overall, sickness was reduced through the visual information generated by the proposed method, supporting the effectiveness of the proposed approach. These findings imply that the reduction in motion sickness experienced by the participants could be attributed to the proposed visual stimulus optimization method by the computational model based on the SVC theory.

Neural mismatch theory or sensory conflict theory [6] suggest that the motion sickness can be reduced when visual and vestibular information are congruent. Various methods have been proposed in the literature to mitigate motion sickness by artificially providing visual information based on this idea when vision is unavailable. For example, research has demonstrated that motion sickness can be mitigated by displaying forward images on a screen to individuals seated in a rear seat [26]. Additionally, there has been an attempt to reduce motion sickness when reading on a smartphone or tablet by presenting bubbles around characters in sync with the vehicle's motion [19]. Additionally, studies have shown that when reading characters with an HMD during vehicle travel, motion sickness is reduced by fixing the characters to the Earth's coordinate system [20] and making the surrounding images visible through video see-through technology [21]. The primary contribution of the present study is the demonstration that a certain type of visual information incongruent with the vestibular system has the potential to reduce motion sickness more effectively than congruent visual-vestibular information.

The present study is part of a growing body of research that utilizes computational models of motion sickness to develop solutions aimed at minimizing the sickness. For example, a previous study focused on the development of vehicle velocity generation methods for AVs during curve negotiation[3]. Additionally, research efforts have been dedicated to the advancement of vehicle motion planners and motion controllers, employing the 6DoF-SVC model [16]. These studies aimed to optimize head and vehicle motions by leveraging vestibular motion sickness models. This study stands as the initial endeavor to optimize visual motion by employing a computational model that integrates dynamics of both the vestibular and visual systems. To strengthen the evidence for the conclusions obtained here, it is necessary to increase the number of participants beyond the initial sample of seven males. Furthermore, the inclusion of female participants is essential to ensure the generalizability of the findings across the sexes. Another limitation of this study is that it focused solely on testing a single-motion paradigm. Consequently, the applicability of the proposed method for various motion patterns remains unknown. Future research is needed to investigate the effectiveness of the proposed method across a wide range of motion patterns to assess its broader applicability. Another limitation is that the head motion of the participants was not considered when generating visual information. In the present study, a neck brace was used to minimize the effects of head movement. It is important to acknowledge that in more realistic driving scenarios, such as situations involving lateral acceleration, accounting for head movement becomes essential to the proposed method. Incorporating head movements into the proposed method increases its applicability under real driving conditions.

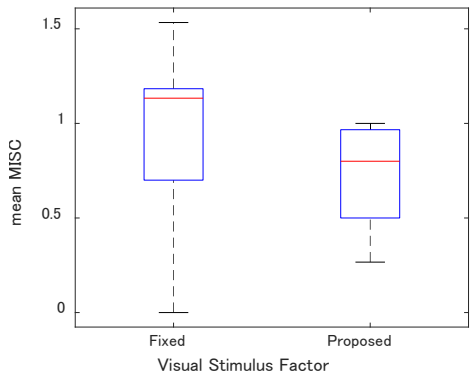

Figure 8. Mean MISC score

Acknowledgment

This work was partially supported by the JSPS KAKENHI (Grant Number 21K18308 and 21H01296), Japan.